\documentclass[a4paper]{article}

\usepackage[dvips]{graphicx}
\usepackage{epsfig}
\usepackage[english]{babel}
\usepackage[ansinew]{inputenc}
\usepackage[T1]{fontenc}
\usepackage{latexsym}
\usepackage{amssymb}

\begin{document}

\title{The density distribution in the Earth along the CERN-Pyh\"asalmi
baseline and its effect on neutrino oscillations}

\author{E. Kozlovskaya\footnote{elena.kozlovskaya@oulu.fi},
J. Peltoniemi\footnote{juha.peltoniemi@oulu.fi}
and J. Sarkamo\footnote{juho.sarkamo@oulu.fi} \\
Center for Underground Physics in Pyh\"asalmi, \\ University of Oulu,
Finland, BOX3000 FIN-90014}

\maketitle

\begin{abstract}
We study the beamline properties of a long baseline
neutrino- oscillation experiment from CERN to the Pyhäsalmi Mine in
Finland.
We obtain the real density profile for this particular neutrino oscillation
beamline by applying the geophysical data. The effects of the matter density to
neutrino oscillations are considered. Also we compare the realistic
density profile with that acquired from the PREM model.
\end{abstract}

\section{Introduction}
A future step towards a better understanding of neutrino properties is to
build long-baseline neutrino-oscillation experiments.
One such possibility
could be to aim a neutrino beam from a proposed CERN Neutrino
Factory\cite{longbaseline} to
the Pyhäsalmi Mine\footnote{http://cupp.oulu.fi} in Finland 2288 km away.
The Neutrino Factory and target station could be operational after the
year 2010.
\par
The oscillation probability depends on the density of the medium as
well as the intrinsic mixing parameters. 
The purpose of this study is to compile a realistic
density profile along the CERN-Pyhäsalmi neutrino baseline and estimate
its effect on the neutrino oscillations. 

\section{Compilation of the density distribution along the baseline}
\subsection{The geographical position of the baseline}

\begin{figure}[ht]
\begin{center}
\includegraphics[width=0.5\textwidth]{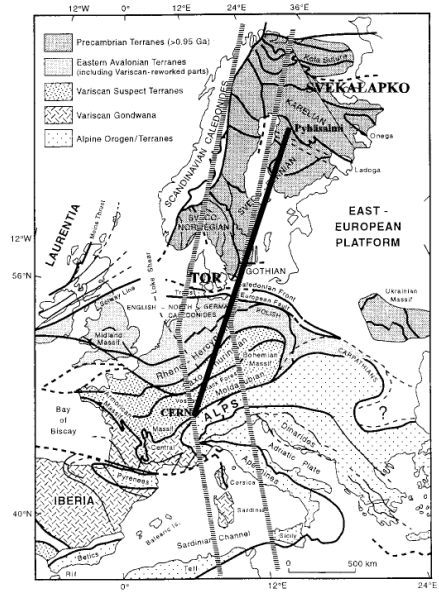}
\caption{Location of the CERN-Pyhäsalmi baseline (solid black line) on the
map
of main tectonic elements of Western Europe\cite{Blundell1992}.
The location of the EGT in the swathe between the two broad lines. }
\label{fig1}
\end{center}
\end{figure}

The neutrino baseline starts 
at $46^\circ$N $15'$ $22.93''$, $06^\circ$E $03'$ $01.10''$, $+450$ m above 
sea level, 
and ends at $63^\circ$N $39'$ $34.74''$, $26^\circ$E $02'$ $29.94''$, $-1345$ 
m. Its total length is $2288$ km. The geographical position 
of the baseline and 
its penetration depth were calculated using the geocentric Cartesian 
coordinate system and
then transformed to the geographical coordinates and depth with
respect to the WGS-84 ellipsoid \cite{Malysh1996} using the transformation
equations in \cite{Bursa1995}. The location of the baseline is
shown on the simplified tectonic map of Europe (Fig. \ref{fig1}).  The baseline
goes mainly through the lithosphere (the Earth's crust
and the uppermost part of the Earth's mantle) and crosses the main geological 
structures and boundaries of the Europe. Its deepest point (103.81 km) is 
located 
below the so-called Trans-European Suture Zone
(TESZ). Thus, the density variations along the line can be due to different thickness of
the crust and different density of the crust and lithospheric mantle
within various tectonic units. Another affecting factor is the depth to the boundary
that separates the non-convecting lithospheric mantle from partially
molten convecting and less dense asthenosphere. This information can be
obtained from results of recent lithospheric studies in Europe.
\begin{figure}
\includegraphics[width=1\textwidth]{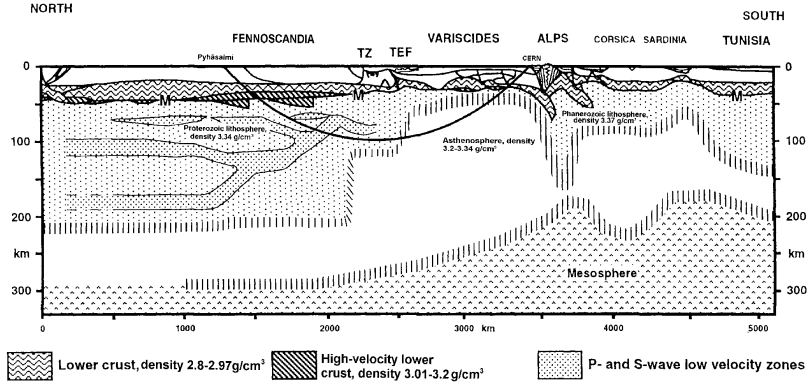}
\caption{
A sketch demonstrating the location of the CERN-Pyhäsalmi baseline on the
composite cross-section of the lithopshere along the EGT
\cite{Blundell1992} (solid black line).
The depth
to the lithosphere-asthenosphere boundary beneath the Trans-European
Suture Zone is modified in accordance with the S-wave velocity model in
\cite{Cotte2002}. M, Moho boundary; TZ Tornquist Zone; TEF,
Trans-European Fault.}
\label{fig2}
\end{figure}
\par
As seen from Fig. \ref{fig1}, the large part of the baseline is
located
within the study area of the European Geotraverse project (EGT). The EGT
was a $4600$ km long and $200-300$ km wide lithospheric transect across
Europe from Norway to Tunisia. The multinational, multidisciplinary
research resulted in a comprehensive cross-section of
the European lithosphere to the depth of $450$ km \cite{Blundell1992}.
It revealed a significant contrast between the thickness 
and structure of the lithosphere of younger western Europe (less than $90$ km) 
and the old cold lithosphere of Fennoscandia and eastern Europe (more than $160$ km),
occurring beneath the TESZ (Fig. \ref{fig2}). Later, the more detailed
information about the lithosphere-asthenosphere boundary beneath the TESZ was obtained
within the seismological EUROPROBE/TOR project
\cite{Artlitt1999etother}\cite{Cotte2002}.
\par
In Finland, the CERN-Pyhäsalmi baseline is located within the study area
of the EUROPROBE/SVEKALAPKO project \cite{HjeltDaly1996etother}.
As a part of the project, the 3D density model of the SVEKALAPKO area on
a $2$ km $\times$ $2$ km $\times$ $2$ km rectangular grid
down to the depth of $70$ km was obtained \cite{Kozlovskaya2002}.
\par
Thus, the abundant geophysical information about structure of the crust and
upper mantle obtained within three large geoscientific projects in Europe
mentioned above (e.g. EGT, TOR, SVEKALAPKO) allows us to compile a
realistic density profile for the CERN-Pyhäsalmi baseline.

\subsection{Composition and density of the continental crust and the
lithospheric mantle}
The thickness of the crust
in Europe varies from $26$ km in its western part to 65 km beneath the
Fennoscandia.  The most abundant elements in the continental crust are O
(46.4 wt$\%$), Si (28.15 wt$\%$), Al (8.23 wt$\%$), Fe (5.63 wt$\%$), Mg
(2.33 wt$\%$), Ca (4.15 wt$\%$), Na (2.36 wt$\%$) and K (2.09 wt$\%$)
\cite{Carmichael1989}, occuring in the Earth in the form of various silicate minerals. 
The density in the continental crust generally increases with depth due to the 
lithostatic pressure resulting in rock compaction and also due to
decrease of the content of SiO$_2$ in the rock-forming minerals of lower crustal 
rocks. The strongest density contrasts in the crust exist between the
sedimentary cover and the bedrock and also at the so-called Mochorovichich
boundary (Moho boundary) separating the crust from the mantle.
\par
The major part of the baseline is located within the upper $100$ km of the
Earth's litospheric mantle. The physical properties and composition of it are known 
from geophysical studies and from direct measurements on samples of mantle rocks 
that have been overthrusted or exhumed to the surface by various tectonic and 
magmatic processes. The main elements in the
upper mantle are Fe, Mg, Si and O. The main rock forming minerals for upper mantle 
rocks (peridotites) are olivine ((Mg,Fe)$_2$SiO$_2$), orthopyroxene ((Mg,Fe)SiO$_2$), 
clinopyroxene ((Ca,Na)(Mg,Fe,Al, Cr,Ti)(Si,Al)$_2$O$_6$) and spinel (MgAl$_2$O$_4$). 
It is believed that the continental 
lithospheric mantle has undergone significant melting
through geological time and is depleted in such components as Fe, Al, Ca
and Ti. As a result, the density of the continental
lithospheric mantle generally depends on the tectonothermal age: younger 
lithospheric mantle is less depleted and hence is denser. The conversion from 
fertile to depleted mantle is expressed by a decrease in clinopyroxene and 
orthopyroxene and relative increase in olivine, and also by an increase of Mg content. 
That is why the  lithospheric mantle beneath the young Phanerosoic Western Europe has 
the average density of $3.37$ g$/$cm$^3$,
while the older (Proterozoic) mantle beneath the Fennoscandian Schield is less
dense ($3.34$ g$/$cm$^3$) \cite{Gaul2002}.
\par
In comparison to the lithospheric mantle, the asthenosphere is fertile and
chemically more homogeneous because of convection. The density of the 
asthenosphere is less than that of the lithospheric mantle and
depends mainly on the density and content of partially molten material 
(tholeitic basalt) \cite{Herzberg1995}.
\par
The amount of partial melt in the asthenosphere can be estimated from
the stability condition for the olivine-molten basalt mixtures. They are  
mechanically stable only in the case when the melt is concentrated within 
isolated, non-connected inclusions(pockets). The theoretical modelling of 
elastic and electrical properties of such mixtures\cite{KozlovskayaHjelt2000}
demonstrated that in some cases only $5\%$ of melt inclusions is enough to
form a perfectly interconnected network. This value (less than $5\%$) 
is in agreement with estimates obtained by teleseismic tomography studies 
\cite{Sobolev1996etother}. Thus, if the melt content is
less than $5\%$, and the density of the molten basalt under upper mantle
pressure-temperature conditions is $2.72$ g$/$cm$^3$ \cite{Manghnani1986}, then the
lithosphere-asthenosphere density contrast is less than $0.04$ g$/$cm$^3$.
\par
The lithosphere-asthenosphere density contrast beneath the TESZ can be 
roughly estimated also from the S-wave velocity ($V_s$)
model in \cite{Cotte2002}. In accordance with it, the
$V_s$ in the mantle lithosphere is $4.57$ km$/$s, while the $V_s$ in the 
asthenosphere is $4.36$ km$/$s. Using the scaling factor relating decrease in $V_s$ to the decrease in density, that is equal to $0.05$ at a depth of about $100$ km \cite{Deschamps2001}, the asthenophere density beneath the TESZ is
$3.34$ g$/$cm$^3$.
\par
These estimates of the asthenosphere density agree with the results of 
regional (medium-wavelength and long-wavelength) gravity studies in
Europe ($3.2-3.33$ g$/$cm$^3$)\cite{MarquartLelgemann1992etother}.
Summarising all the data about the density in the asthenosphere cited
above, we can conclude that the range of possible values of the
asthenosphere density for Western Europe is $3.2-3.34$ g$/$cm$^3$.

\subsection{Geological setting and density values along the baseline}
From $0$ to $175$ km the baseline goes through the Earth
crust beneath the Jura Mountains, Molasse Basin and the Upper Rhine Graben.
The thickness of the crust here is $30-28$ km \cite{GieseBuness1992}.
The first $21$ km of the line are located within the sedimentary cover that is 
$3300-3400$ m thick\cite{Sommaruga1999}. The density within the 
sedimentary cover composed of limestones, sandstones and shales
gradually increases with depth from $2.4$ to $2.6$ g$/$cm$^3$.
\par
The next portion of the line ($22-175$ km) goes through the upper($22-71$ km),
middle ($72-119$ km) and lower crystalline crust ($120-175$ km) with the densities 
of $2.7-2.73$ g$/$cm$^3$, $2.8-2.87$ g$/$cm$^3$ and
$2.96-2.97$ g$/$cm$^3$, respectively. The thickness of the layers and their density were taken from 
the 3-D gravity studies\cite{Ebbing2002}\cite{Gutscher1995}.
\par
The next part of the baseline ($175-732$ km) goes through the Palaeosoic
lithospheric mantle ($3.37$ g$/$cm$^3$). Basing on 
studies by \cite{Gutscher1995}\cite{AchauerMasson2002} we
assumed no upwelling asthenosphere beneath the Upper Rhine Graben ($175-430$ km). 
Thus, the baseline goes through the uprising asthenosphere only at $732-1019$ km.
As it was shown in previous Chapter, the density of the asthenosphere here can be $3.20-3.34$
g$/$cm$^3$. The lithosphere-asthenosphere boundary along the baseline was estimated in
accordance with the lithosphere thickness map\cite{BabushkaPlomerova1992} 
and from the model\cite{Cotte2002}, with the accuracy of $50$ km.
\par
From $1020$ to $1956$ km the baseline is located within the Proterozoic lithospheric
mantle ($3.34$ g$/$cm$^3$), then it
returns to the crust at a depth of $52.01$ km. The density values along the final
part of the baseline ($1957-2288$ km) were taken from the 3-D density model
of the SVEKALAPKO area\cite{Kozlovskaya2002}. The crust in this part of Finland
consists of four major layers: the upper crust($2.73-2.81$ g$/$cm$^3$),
the middle crust($2.86-2.89$ g$/$cm$^3$), the lower crust($2.86-2.89$ 
g$/$cm$^3$) and so-called high-velocity lower crust($3.01-3.2$ g$/$cm$^3$).
The density values along the baseline are summarised in Figure
\ref{rhoos.ps}.
\section{Neutrino matter oscillations along the baseline}
Next we will discuss the effect of the realistic CERN-Pyh\"asalmi density
profile to neutrino-oscillations. In a general three flavour framework 
the distance evolution equation of the neutrino flavour states can be 
written as\cite{mattereffect}
\begin{equation}
i {\textrm{d} \over \textrm{d}x} \pmatrix{\nu_e \cr \nu_\mu \cr \nu_\tau} =
\left({1\over 2E} U 
{M}^2
U^\dagger  + \pmatrix{\Delta V & 0 & 0 \cr 0 & 0& 0 \cr 0& 0&
0}\right)
\times \pmatrix{\nu_e
\cr \nu_\mu \cr \nu_\tau}
\label{eq1}
\end{equation}
and we use the PDG endorsed parametrization\cite{PDGU}
for the unitary mixing matrix $U$. 
The potential term $\Delta V = \pm\sqrt{2}G_F 
Y_e \rho$ arises from the coherent forward charged-current scatterings of
electron neutrinos from the matter electrons. Throughout this paper we
assume $Y_e = 0.5$, because of the crust and mantle material content,
which is mainly Si and O.
We solve the equation (\ref{eq1}) numerically by using a fourth order Runge-Kutta -algorithm.
\par
\begin{figure}[ht]
\begin{center}
\includegraphics[angle=0, width=0.75\textwidth]{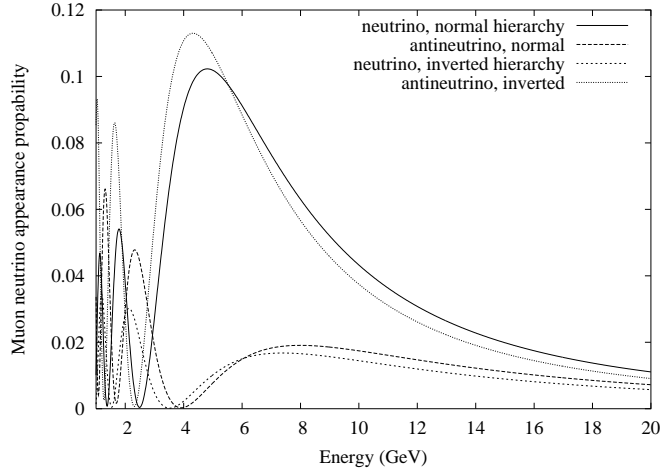}
\caption{An example of muon neutrino and antineutrino appearance
probabilities for different energies and mass hierarchies.
(Parameters set to
$\delta m_{21}^2 = 5\cdot 10^{-5} \textrm{eV}^2$, $\delta m_{32}^2
=\pm 3.15\cdot 10^{-3}\textrm{eV}^2$, $\sin^2 2\theta_{12}=0.87$,
$\sin^2 2\theta_{23}=1.0$, $\sin^2 2\theta_{13}=0.1$, $\delta=0$)
}
\label{kuva.ps}
\end{center}
\end{figure}

At energies relevant to a neutrino factory beam (above $1$ GeV),
the $\nu_e \rightarrow \nu_{\mu}$ -oscillations are governed mainly by
the mixing angle $\theta_{13}$ and the mass squared difference $\delta
m_{31}^2\approx \delta m_{32}^2$. For typical values of $\delta m_{31}^2 
= +3.0\cdot 10^{-3}\textrm{ eV}^2$, $\theta_{13} = 0.1$ 
and density $\rho \approx 3.3$ g cm$^{-3}$, the matter
resonance energy in the leading approximation\cite{Geer}
is about $\sim 11$ GeV, but the distance $2288$ km is much smaller than
the corresponding oscillation lenght ($L_{osc} \sim 45000$ km) and
therefore no total resonance conversion can happen. 
Our simulations show that
the muon neutrino appearance probability at the first oscillation maximum
${E \sim 5}$ GeV is
enhanced by about a factor of two compared to the appearance probability
in vacuum. For a negative sign of $\delta m_{31}^2$, corresponding
to the inverted mass hierarchy, the appearance probability is
suppressed by about a factor of two.
For $\overline{\nu}_e \rightarrow \overline{\nu}_{\mu}$-oscillations the
situation is opposite (shown in Figure \ref{kuva.ps}).
\par

\begin{figure}[p]
\begin{center}
\includegraphics[angle=0, width=0.7\textwidth]{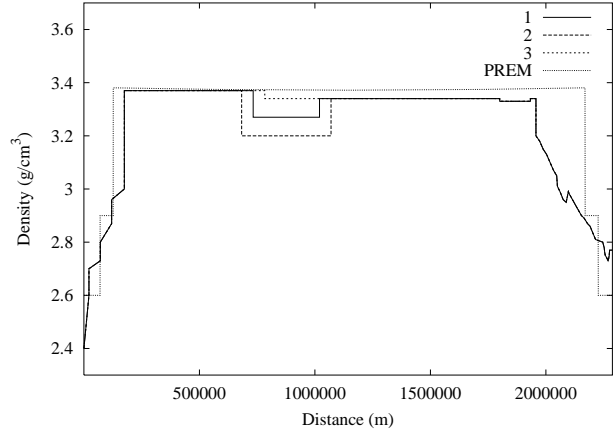}
\caption{The different density profiles used for error analysis.}
\label{rhoos.ps}
\end{center}
\end{figure}
\begin{figure}[p]
\begin{center}
\includegraphics[angle=0,width=0.7\textwidth]{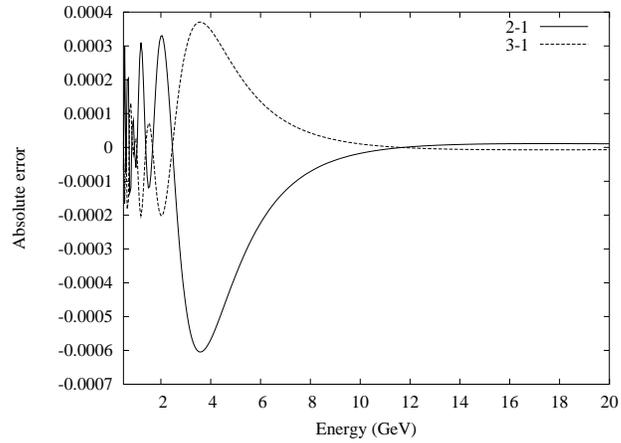}
\caption{The effect of the astenospheric uncertainties to the muon
neutrino appearance probability. Absolute differences of matter profiles 2 
and 3 from the density profile 1 as a function of energy.
(Parameters set as in Figure \ref{kuva.ps}, with $\delta m_{32}^2>0$.)
}
\label{Absoluteerror.ps}
\end{center}
\end{figure}

The largest uncertainty in the obtained density profile is the
astenosphere at $L = 733\dots 1010$ km, whose size and density can
vary by $\Delta L = \pm 100$ km and $\rho = 3.2\dots3.34$ g cm$^{-3}$. 
Because of this uncertainty, we consider three possible density
profiles (shown in Figure \ref{rhoos.ps} together with the profile acquired 
from the PREM-model\cite{PREM}): an average density and average astenosphere 
width profile with $\rho_{ast}=3.27$ g cm$^{-3}$
(1), a wide astenosphere with low density  $\rho_{ast}=3.20$ g cm$^{-3}$ (2) 
and a narrow astenosphere with high density  $\rho_{ast}=3.34$ g
cm$^{-3}$ (3). 
We compare the muon neutrino appearance probabilities calculated with
the different matter 
profiles. The absolute differences are shown in Figure
\ref{Absoluteerror.ps}. We see that, for an upper limit value\cite{CHOOZ}
$\sin^2 2\theta_{13}=0.1$,
the differences between the appearance probabilites at oscillation maximae
are of the order of few $10^{-4}$. 
This corresponds to a relative error below $1\%$.
We conclude that the errors of the uncertain nature of the astenosphere
can be considered as a small, often negligible, error in future analysis.

\section{Discussion}
The PREM density profile is commonly used in Earth matter density neutrino
oscillation studies. There has been discussion\cite{disc} about the
accuracy of this approximate model.
Therefore we also study the oscillation probability differences between
the PREM and the realistic density profile.
\par
\begin{figure}[Ht]
\begin{center}
\includegraphics[angle=0,width=0.7\textwidth]{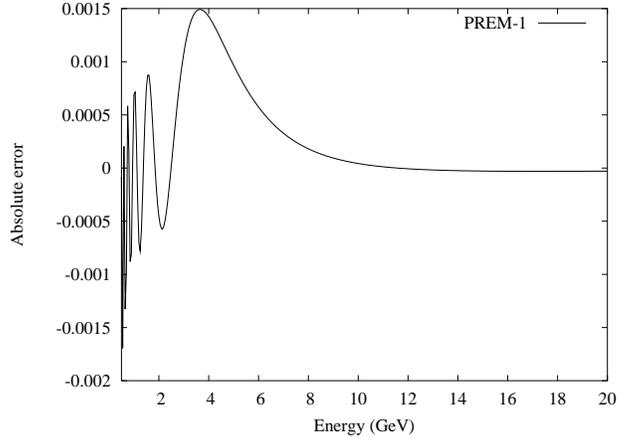}
\caption{The absolute difference of the muon neutrino appearance
probabilities $P(\textrm{PREM})-P(\textrm{REAL})$. (Parameters set as in
Figure \ref{Absoluteerror.ps}.)}
\label{PREMabsoluteerror.ps}
\end{center}
\end{figure}

We consider the $\nu_e \rightarrow \nu_{\mu}$ -oscillations. The 
absolute difference for the muon neutrino appearance probability between
the two density profiles is shown in Figure \ref{PREMabsoluteerror.ps}. 
The difference of the appearance probability
between the two density profiles is, around the first oscillation maximum, 
about a factor of 3 times larger than the error due to the
astenospheric uncertainties. So, in
obtaining the realistic density profile, the errors of
neutrino oscillation simulations are a small step better, allthough they
are not drastically improved.
\par
Random matter fluctuations of the density profile have been
considered in Ref. \cite{Ohlsson}. It is estimated
that with different realistical 
random matter fluctuations the average error of $P_{\mu e}$ for a distance
$L \approx 2300$, average density $\rho = 3$ g cm$^{-3}$ and energy
$E_{\nu} = 30$ GeV is $|\Delta P_{\mu e} | \sim
0.5 \dots 1 \cdot 10^{-5}$. 
In our case (with parameters as in Ref.\cite{Ohlsson}), the
difference between the PREM and the realistic case is
$|\Delta P_{\mu e}| = 0.4\cdot 10^{-5}$ and $|\Delta P_{{\overline
\mu}{\overline
e}}| = 0.6\cdot 10^{-5}$. Hence it seems that the error result from the
random matter fluctuation model fits nicely in to our 2288 km baseline.
\par
We will also like to point out that the calculated relative error
between the neutrino factory beam muon event rates (from
$\nu_e \rightarrow \nu_{\mu})$ using the PREM and
realistic density profiles is below $\sim 2\%$ for
beam energies around
$1-10$ GeV and is considerably smaller for larger energies.
\par
We conclude that we have made up a
reference model for the CERN- Pyhäsalmi baseline that is accurate enough
for all studies of long baseline
neutrino oscillations. Although the results are quite similar to the PREM
model, there is no excuse for using a less
accurate model. However, for other baselines that have not been
specifially modelled the PREM model may be a good approximation to start
with.
\par
The calculated event rates and parameter reaches of the CERN-Pyhäsalmi 
neutrino beam will be discussed in a future paper.

\section*{Acknowledgements}
This work has been partially funded by EU structural funds.


\begin{thebibliography}{99}
\bibitem{longbaseline} M. Apollonio et al., CERN-TH/2002-208
\bibitem{Malysh1996} S. Malys, The WGS84 Reference Frame. National
Imagery and mapping agency, Nov. 7, (1996).
\bibitem{Bursa1995} M. Bursa, Earth, Moon and Planets, 69, 51 (1995)
\bibitem{Blundell1992}
D.J. Blundell, R. Freeman, St. Mueller, St. (ed) A continent revealed: the European Geotraverse 1992.
Cambridge University Press, Cambridge (1992)

\bibitem{Artlitt1999etother}
R. Artlitt,  PhD Thesis, ETH, Zurich (1999)\\
S. Gregersen et al., Tectonophysics, 360, 61 (2002)\\
J. Plomerova et al., Tectonophysics, 360, 89 (2002)
\bibitem{Cotte2002}
N. Cotte et al., Tectonophysics 360, 75 (2002)
\bibitem{HjeltDaly1996etother}
S.-E. Hjelt and S. Daly et al., In D.G. Gee and H.J. Zeyen (ed) EUROPROBE 1996 - Lithospheric
dynamics: Origin and evolution of continents. Uppsala University (1996)\\
G. Bock (ed) et al., EOS Transactions AGU, 82, 50, 621, 628-629 (2001)

\bibitem{Kozlovskaya2002}
E. Kozlovskaya et al. Geoph. J. Int., submitted (2003).
\bibitem{Carmichael1989}
R. Carmichael (ed) Practical handbook of
physical properties of rocks and minerals. CRC Press, Boca Raton, Florida (1989)
\bibitem{Gaul2002} O.F. Gaul et al.,  Earth Plan. Sci. Lett., 182, 223 (2000).
\bibitem{Herzberg1995} C. Herzberg. Phase equilibra of common rocks
in the crust and mantle. In: Ahrens, T. (ed) Rock physics and phase relations. A handbook
of physical constants. AGU (1995).
\bibitem{KozlovskayaHjelt2000}
E. Kozlovskaya and S.-E. Hjelt, Phys. and Chem. of the Earth (A),25, 2, 195 (2000)

\bibitem{Sobolev1996etother}
S. Sobolev et al., Tectonophysics, 275, 143 (1997)\\
H. Sato et al., Pure Appl.Geoph., 153, 377 (1998)\\
C. Petit et al., Earth Plan. Sci. Lett.,197, 3-4, 133 (2002)

\bibitem{Manghnani1986} M. Manghnani et al., J. Geoph. Res, 91, B9, 9333 (1996)
\bibitem{Deschamps2001} F. Deschamps et al., Phys. Earth Plan. Int., 124, 193 (2001)

\bibitem{MarquartLelgemann1992etother} G. Marquart and D. Lelgemann, Tectonophysics, 207,
25 (1992)\\
F. Cella et al., Tectonophysics, 287, 1-4, 117 (1998)

\bibitem{GieseBuness1992} P. Giese and H. Buness. Moho depth. In:
D.J. Blundell, R. Freeman, St. Mueller (ed) A continent revealed: the European
Geotraverse 1992. Cambridge University Press, Cambridge (1992)
\bibitem{Sommaruga1999}
A. Sommaruga, Marine and Petroleum Geology, 16, 111 (1999)
\bibitem{Ebbing2002} J. Ebbing, PhD Thesis, Fachbereich
Geowissenschaften, Freie Universität Berlin (2002)
\bibitem{Gutscher1995} M.-A. Gutscher, Geophys. J. Int., 122, 617 (1995)
\bibitem{AchauerMasson2002}
U. Achauer and F. Masson, Tectonophysics, 358, 17 (2002)
\bibitem{BabushkaPlomerova1992}
V. Babushka and J. Plomerova, Tectonophysics, 207,141 (1992).

\bibitem{mattereffect}L. Wolfenstein, Phys. Rev. D17, 2369 (1978)
\bibitem{PDGU} K. Hagiwara et al., Phys. Rev. D66, 010001 (2002) 
\bibitem{Geer} V. Barger et al., Phys. Rev. D62, 013004 (2000)
\bibitem{PREM} A. M. Dziewonski and D.L. Anderson, Phys. Earth Planet.
Inter. 25, 297 (1981), 
\bibitem{CHOOZ} M. Apollonio et. al, Eur. Phys. J. C27, 331 (2003)
\bibitem{disc} for example, see R. Geller and T. Hara, hep-ph/0111342
(2001)
\bibitem{Ohlsson} B. Jacobsson et al., Phys. Lett. B532, 259 (2002)
\end{thebibliography}
\end{document}